\documentclass[floatfix,prb,twocolumn,superscriptaddress]{revtex4-2}
\usepackage[utf8]{inputenc}
\usepackage{booktabs}
\usepackage{geometry}
\newgeometry{vmargin={13mm}, hmargin={12mm,16mm}}   
\usepackage{amsmath,amssymb,stmaryrd}   
\usepackage{xcolor}
\usepackage{amsmath}
\usepackage{upgreek}
\usepackage{parskip}
\usepackage{xfrac} 
\usepackage{graphicx}
\usepackage{lipsum}
\usepackage{dcolumn}
\usepackage{bm}
\usepackage{siunitx}
\usepackage{float}
\usepackage{esvect}
\usepackage{gensymb}
\usepackage{commath}
\usepackage{mathtools}  
\usepackage{diffcoeff}
\usepackage[version=4]{mhchem}
\usepackage{chemformula}
\usepackage[colorlinks=true,linkcolor=blue]{hyperref}%

\begin{document}

\preprint{APS/123-QED}

\title{Strain control of band topology and surface states in  antiferromagnetic \ch{EuCd2As2}}

\author{Nayra A. \'Alvarez Pari}
\affiliation{Institut für Physik, Johannes Gutenberg Universität, D-55099 Mainz, Germany}
\author{V. K.  Bharadwaj}
\affiliation{Institut für Physik, Johannes Gutenberg Universität, D-55099 Mainz, Germany}
\author{R.  Jaeschke-Ubiergo}
\affiliation{Institut für Physik, Johannes Gutenberg Universität, D-55099 Mainz, Germany}
\author{A.  Valadkhani}
\affiliation{Institut für Theoretische Physik, Goethe-Universität Frankfurt, 60438 Frankfurt am Main, Germany}
\author{Roser  Valentí}
\affiliation{Institut für Theoretische Physik, Goethe-Universität Frankfurt, 60438 Frankfurt am Main, Germany}
\author{L.  \v{S}mejkal}
\affiliation{Institut für Physik, Johannes Gutenberg Universität, D-55099 Mainz, Germany}
\affiliation{Inst. of Physics Academy of Sciences of the Czech Republic, Cukrovarnick\'{a} 10,  Praha 6, Czech Republic}
\author{Jairo Sinova}
\affiliation{Institut für Physik, Johannes Gutenberg Universität, D-55099 Mainz, Germany}
\affiliation{Department of Physics, Texas A\&M University, College Station, Texas 77843-4242, USA}
\date{\today}

\begin{abstract}
Topological semimetal antiferromagnets provide a rich 
source of exotic topological states which can be controlled by manipulating the orientation of the Néel vector, or by modulating the lattice parameters through strain. We investigate via {\it ab initio} density functional theory calculations, the effects of shear strain on the bulk and surface states in two antiferromagnetic  \ce{EuCd2As2}
phases with out-of-plane and in-plane spin configurations. When magnetic moments are along the \textit{c}-axis,  a 3\% longitudinal or diagonal shear strain can tune the Dirac semimetal phase to an axion insulator phase, characterized   by the parity-based invariant $\eta_{4I} = 2$. For an in-plane magnetic order, the axion insulator phase  remains robust under all shear strains. We further find that for both magnetic orders,  the bulk gap increases and a surface gap opens on the (001) surface up to 16 meV. Because of a nonzero $\eta_{4I}$ index and  gapped states on the (001) surface, hinge modes are expected to happen on the side surface states between those gapped surface states. This result can provide a valuable insight in the realization of the long-sought axion states.
\end{abstract}
\maketitle

\section{Introduction}

The discovery of topological magnetic materials \cite{xu2020high, otrokov2019prediction,li2019dirac, zhang2019topological,zou2019study, hua2018dirac} has generated substantial interest from the scientific community, primarily due to their potential for new applications of quantum materials. These magnetic compounds possess unique characteristics  for achieving dissipationless spin and charge transport, creating chiral electronic channels with perfect conduction and enabling energy-efficient spintronics, among others \cite{bernevig2022progress, vsmejkal2018topological, fan2016spintronics}. Due to the interplay between topological properties 
and  magnetic order,  diverse topological phases can emerge \cite{zheng2017field, li2020intrinsic, zheng2015magnetic}.  

\begin{figure}[H]
	\includegraphics[width=0.5\textwidth]{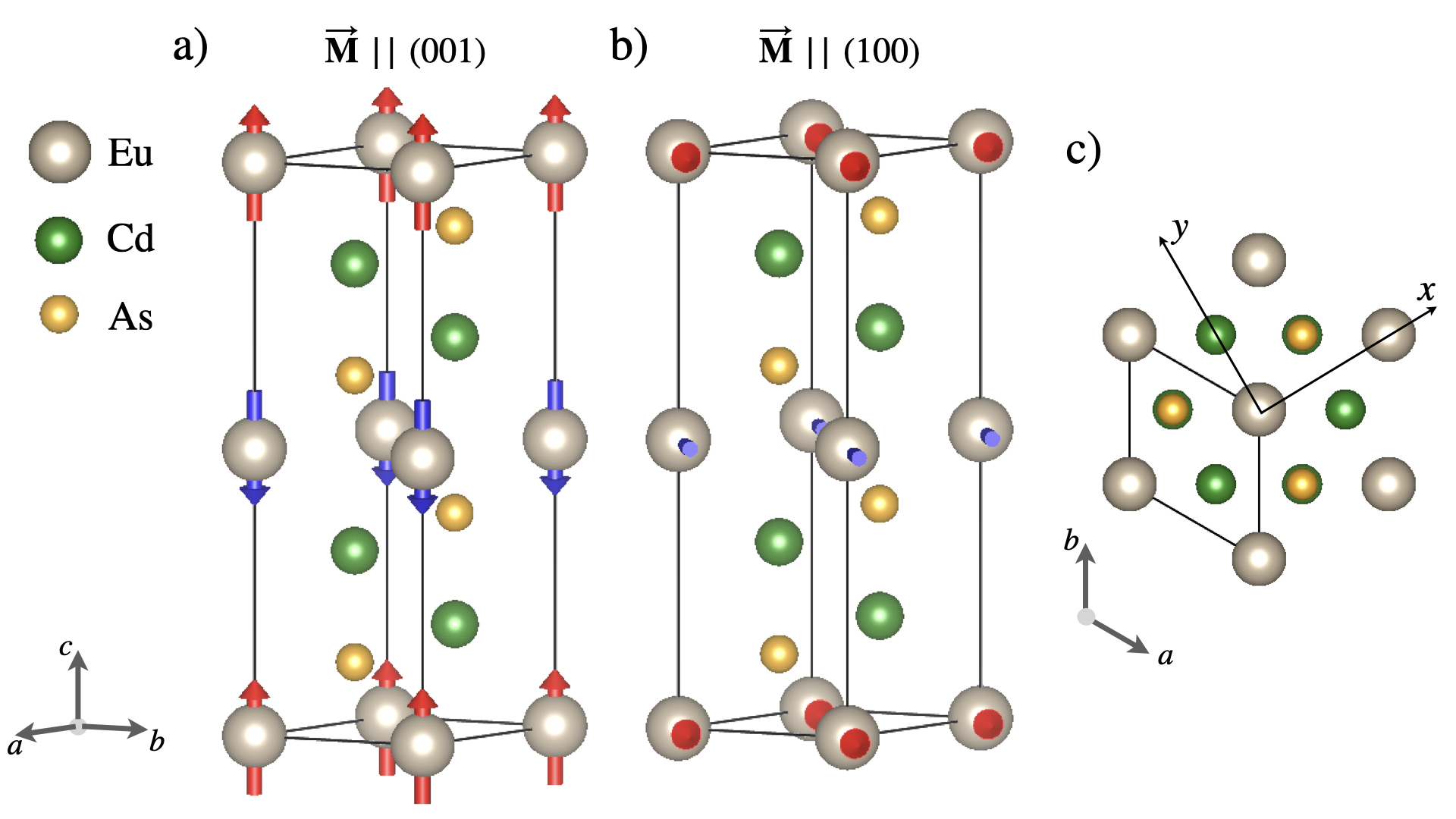}
	\caption{Magnetic unit cell of the two most stable interlayer AFM phases of \ce{EuCd2As2}. a) \textit{afm-oop}: Out-of-plane N\'eel vector orientation along the c-axis, (001). b)    \textit{afm-ip}: In-plane N\'eel-vector orientation along (100). Here (001) and (100) are in terms of the Cartesian coordinates. c) Top view of the hexagonal lattice.}
 \label{struct}
\end{figure}

Antiferromagnetic Dirac semimetals (DSMs) provide a suitable platform for exploring  exotic topological phases by breaking specific symmetries through the manipulation of the N\'eel vector orientation or  application of strain \cite{liu2014stable, aggarwal2016unconventional, neupane2014observation, hua2018dirac, ma2020emergence}. They host quasi-particle excitations near four-fold degenerate band crossings, commonly called Dirac points (DPs). Implementing the density functional theory (DFT) within the standard generalized gradient approximation (GGA) +Hubbard U set to 5 eV, \ce{EuCd2As2} has been identified as an AFM DSM candidate material hosting a single pair of DPs \cite{cuono2023}. The degeneracy of these Dirac fermions is protected by a combined symmetry $\mathcal{PT'}$, where $\mathcal{P}$ is inversion, and  $\mathcal{T'}=\mathcal{T} \oplus \tau$ is a nonsymmorphic time-reversal symmetry, with $\tau=(0,0,c/2)$ a translation operator that connects the up and down spin momentum layers.   This material also exhibits a small magnetic anisotropy energy (MAE) between its two most stable interlayer antiferromagnetic states \cite{hua2018dirac,rahn2018coupling}, with \ce{Eu} moments pointing out-of-plane (\textit{afm-oop}), or along the in-plane (\textit{afm-ip}), as indicated in  Fig. \ref{struct}. A  small MAE suggests that it is possible to manipulate the N\'eel vector orientation.


\begin{figure*}
	\includegraphics[width=\textwidth]{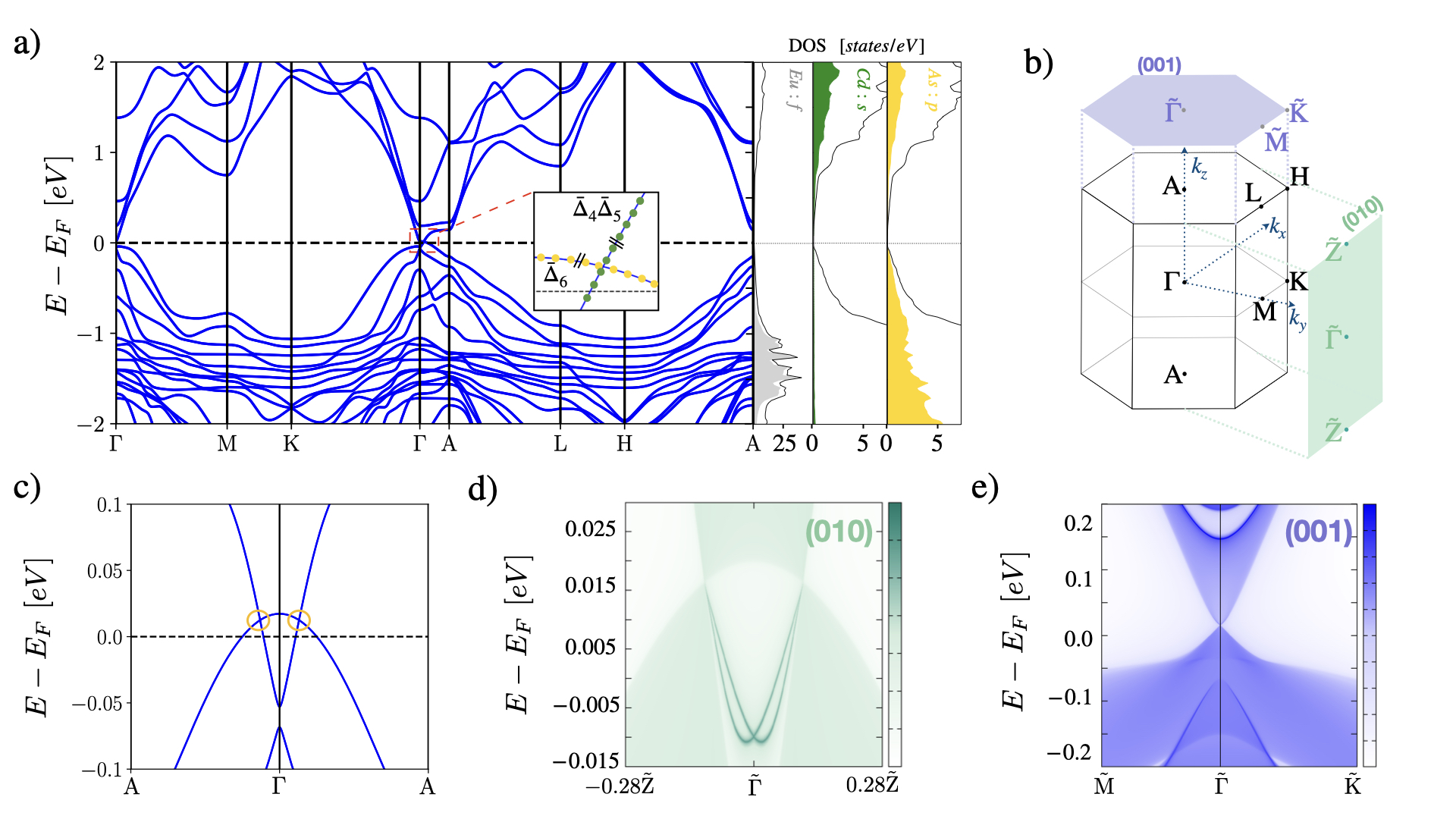}
	\caption{Results for the AFM \ce{EuCd2As2} state with out-of-plane N\'eel vector orientation. a) Electronic band structure
  GGA+SOC+U along the high-symmetry path of the hexagonal lattice with Hubbard U = 5 eV and the density of states of Eu 4f (gray), Cd 5s (green) and As 4p (yellow) orbitals. The black lines display the total DOS.
  b) Bulk First Brillouin zone (BZ) of the $P\bar{3}m1$ hexagonal space group with high symmetry points and the surface BZ projected onto (001) and (010) surfaces. Here (001) and (010) indicate the surface normal vector in terms of the Cartesian coordinates. c) Pair of DPs along the $C_3$ fold rotation symmetry axis. d) Surface states on the $k_y = 0$ plane (010) surface along the Z-$\Gamma$-Z path. Here it is zoomed-in up to the distance of 0.28 between the $\Gamma$ and $Z$ point. e) Surface states along the M-$\Gamma$-K path on the $k_z = 0$ plane (001) surface.}
 \label{oop}
\end{figure*}


In a recent work~\cite{valadkhani2023influence} we analyzed the interplay of Eu magnetism, strain and pressure on the realization of topological phases in \ce{EuCd2As2} by a
combination of a group theoretical analysis with {\it ab initio} density functional theory calculations and found a rich phase diagram with various non-trivial topological phases beyond a Weyl semimetallic
state, such as axion and topological crystalline insulating phases. 
In the present work, we focus on the effect of shear strain.
For a  3\% lattice modulation under shear strain along three in-plane directions, we observe the tuning from the DSM phase to the axion insulating  phase within an \textit{afm-oop} order. When an  \textit{afm-ip} order is considered, the axion insulating phase remains unaltered.
Additionally, for both magnetic orders we observe a gap opening in the surface states on the top and bottom (001) surfaces, which together with the parity-based invariant $\eta_{4I} = 2$, allows for the emergence of hinge states along the edges of the (001) surface.

This article is organized as follows. Section \ref{sectionII} describes the computational methods employed for our calculations. Section \ref{sectionIIIa} and  \ref{sectionIIIb} present  the electronic band structures and the surface states under longitudinal and diagonal shear strains with an \textit{afm-oop} and an \textit{afm-ip} magnetic order, respectively. For both magnetic orders, we provide the symmetry-based indicators for the bulk and a symmetry analysis to describe the gapped and gapless surface states that emerge on specific surfaces. We also discuss the possibility to realize chiral hinge modes, which can be detected by transport measurements. Finally, in section \ref{sectionIV}, we present our conclusions.

\begin{figure*}
	\includegraphics[width=\textwidth]{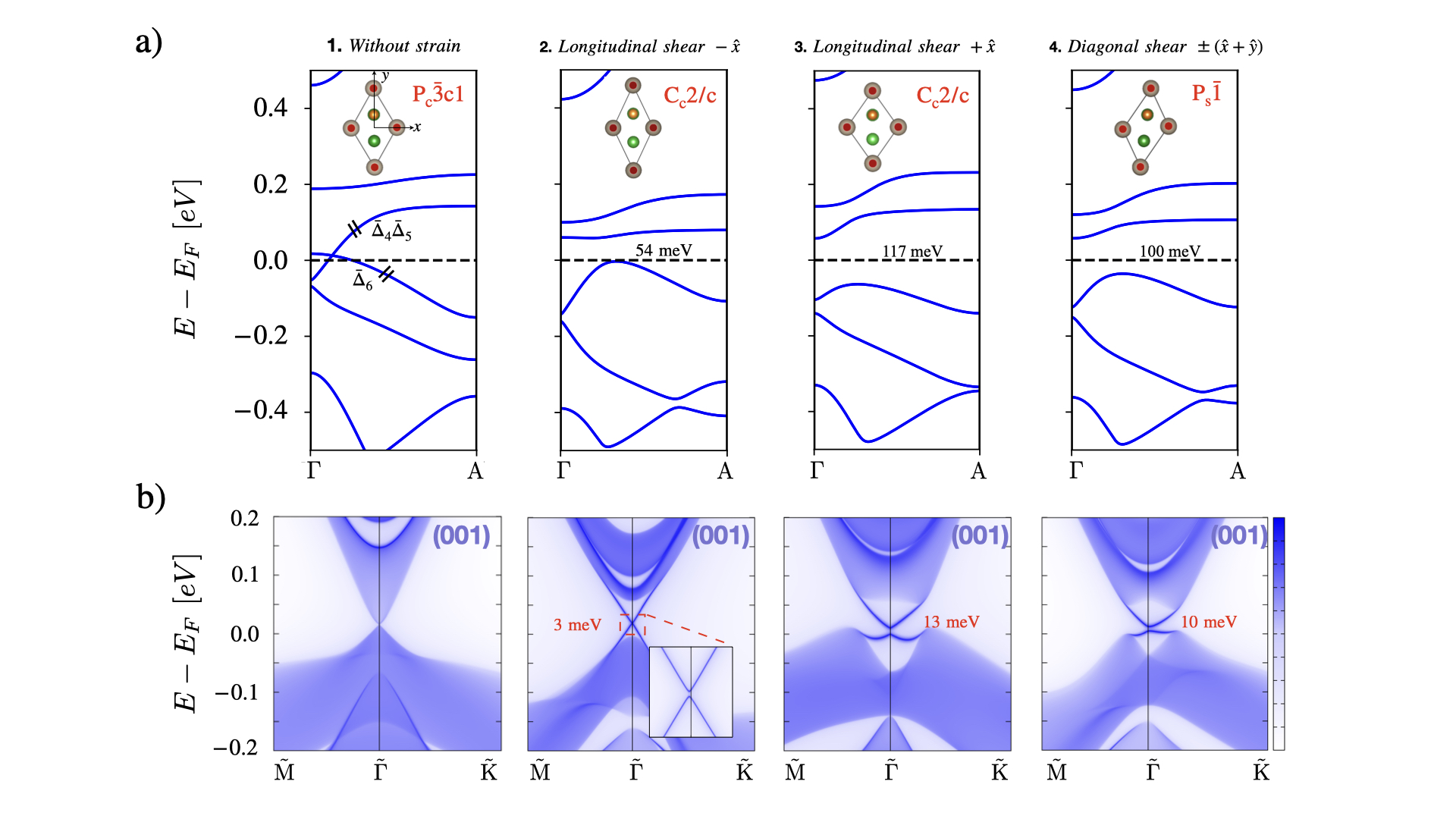}
	\caption{Longitudinal and shear strain application to the AFM \ce{EuCd2As2} state with out-of-plane N\'eel vector orientation. a) GGA+SOC+U energy bands for untrained and strained structures. The top structures aid in visualizing the type of strain and the deformation of the unit cell. The Magnetic Space group (MSG) is given at the top right. The band crossing near the Fermi energy in the unstrained structure (left panel) is a consequence of  $C_3$ symmetry conservation since these bands correspond to different irreducible corepresentations, namely $\bar{\Delta}_{6}$ and $\bar{\Delta}_{4}\bar{\Delta}_{5}$. The applied strain breaks $C_3$ symmetry and a gap opens (middle and right
 panels).
 b) Surface states for unstrained (left panel) and strained (middle and right panels)
 structures along the M-$\Gamma$-K path on the (001) face.}
 \label{strainoop}
\end{figure*}

\section{Computational methods}
\label{sectionII}

The bulk electronic structure calculations and the crystal structure relaxations
under shear strain were performed with the Vienna ab-initio Simulation Package (VASP) \cite{hafner2008ab}. The calculations were conducted using the Perdew-Burke-Ernzerhof (PBE)  exchange-correlation functional within the generalized gradient approximation (GGA) framework \cite{perdew1998perdew}. Spin-orbit coupling was taken into account in all calculations. Additionally, due to localized nature of the Eu 4f orbitals, the GGA+U approach was employed with a value of U set to 5 eV.
A discussion on the use of different exchange correlation functionals can be
found in~\cite{cuono2023}.
To sample the Brillouin zone, a $\Gamma$-centered k-point mesh of $22\times22\times6$ was employed, and the plane wave basis set was extended up to a kinetic energy cutoff of 550 eV. Both of these parameters were carefully checked to ensure convergence.

The structure relaxation was performed by considering the experimental lattice parameters reported by Schellenberg {\it et al.}~\cite{schellenberg2011121sb} until the total force acting on each atom was less than 0.001 eV/Å.  

To determine the topological magnetic indices, we use the implementation of Magnetic Topological Quantum Chemistry (MTQC) \cite{elcoro2021magnetic}. To obtain  the parity eigenvalues and irreducible representations (irreps), we make use  of \textit{Mvasp2trace} \cite{vergniory2019complete} and the Bilbao Crystallographic Server. Furthermore, we employ the Wannier90 package to generate the Wannier Hamiltonians based on projected Wannier functions for $p$ orbitals of As atoms and $s$ orbitals of Cd atoms. The Wannsymm package \cite{zhi2022wannsymm} is also employed to symmetrize the Wannier Hamiltonians and restore the symmetries that can be broken during the disentanglement steps. Finally, to compute the surface states and mirror Chern numbers, the WannierTools \cite{wu2018wanniertools} package is employed.


\section{Results and discussions}
\label{sectionIII}

\subsection{Shear strain effect on the \textit{afm-oop} \ch{EuCd2As2} state}
\label{sectionIIIa}

Without strain or pressure effects,  the symmetry of \ch{EuCd2As2}  with an \textit{afm-oop} order is described by the magnetic space group \textbf{$P_c\bar{3}c1$} (No. 165.96). Figure \ref{oop}(a) shows the GGA+SOC+U electronic band structure along a high symmetry path of the hexagonal 1BZ. The bands near the Fermi level are primarily of  As $p$  and Cd $s$ orbital character, while  Eu 4$f$ states are localized
in the region around -1.5eV forming almost flat bands, as shown in the density of states (DOS) in Fig. \ref{oop}(a).  

The \textit{afm-oop} magnetic order hosts a pair of Dirac points located slightly above the Fermi level, along the $\mathrm{\Delta = \Gamma-A}$ high symmetry line or $C_3$ fold rotation axis Fig. \ref{oop}(d). The Dirac
points originate from
the crossing of  Cd $s$ and As $p$ bands  since  they correspond to different irreducible corepresentations, namely $\bar{\Delta}_{6}$ and $\bar{\Delta}_4 \bar{\Delta}_5$ \cite{valadkhani2023influence}
(see inset in Fig. \ref{oop}(a)).  According to the topological analysis based on  MTQC, the \textit{afm-oop} order is then classified as an enforced semimetal, suggesting that the bulk manifests the characteristics of a topological semimetal like the well known 3D DSM \ce{Na3Bi} \cite{liu2014discovery}.
The bulk DPs are connected by Fermi Arcs, which can be observed through the surface states on the (010) face, as shown in Fig. \ref{oop}(e).  The surface states on the (001) face shows a primarily contribution from the bulk, exhibiting a bulk band touching at the $\bar{\Gamma}$ point, as shown in  Fig. \ref{oop}(f).

\begin{figure*}
	\centering
	\includegraphics[width=\textwidth]{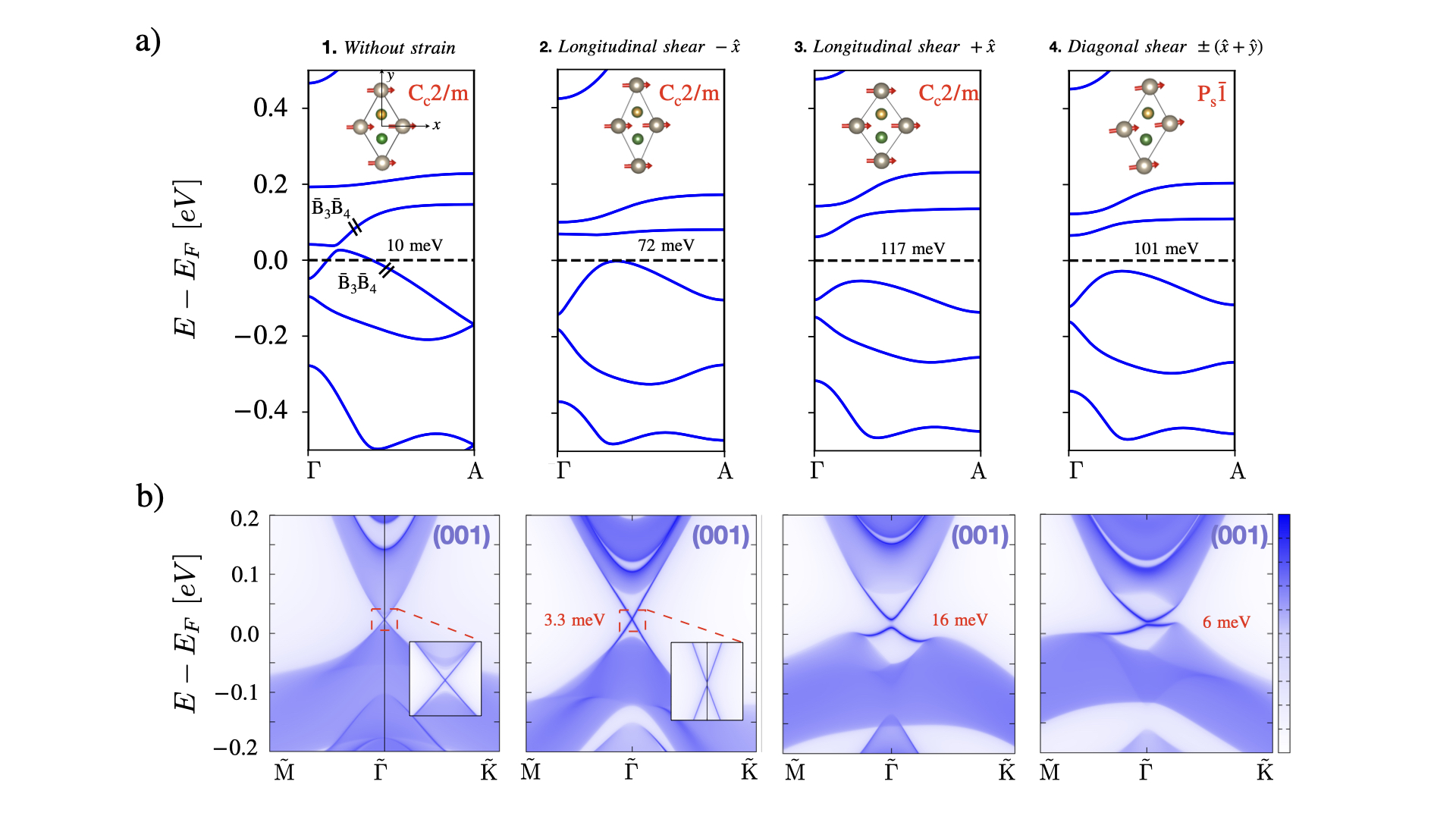}
	\caption{Longitudinal and shear strain application to the AFM \ce{EuCd2As2} state with in-plane N\'eel vector orientation. a) GGA+SOC+U energy bands for unstrained (left panel) and strained (middle and right panels) structures. The top inset structures aid in visualizing the deformation of the unit cell under the given strain. The magnetic space group (MSG) is given at the top right. b) Corresponding surface states for the unstrained and strained structures
  along the M-$\Gamma$-K path on the (001) face.}
 \label{ip-strain}
\end{figure*}

Fig. \ref{strainoop} shows the bulk bands and the surface states on the (001) face for the unstrained
and strained structures  along three different directions on the \textit{ab} plane. We induce a 3\% change in the lattice parameters for all the cases. This modulation leads to a relatively small deformation of the Brillouin zone. Under longitudinal shear strain, both positive (along $+x$) and negative (along $-x$) strain modulations lower the magnetic symmetry group of the system to $\mathrm{C_c2/c}$ MSG (No. 15.90), whose generators are $\mathcal{T'}$, inversion $\mathcal{I}$, and a two-fold rotation with a translation along the z-axis $C_{2x}'= \{C_{2x}|00c/2\}$. Both deformations not only enhance the contribution from the surface states but also open a gap up to 13 meV. This gap opening takes place due to the breaking of  $\mathcal{T'}$ symmetry and the glide mirror $M_x' =\{M_x|00c/2\}$ when open boundary conditions are imposed along the c-axis for the calculation of the surface states. 
On the other hand, the application of  diagonal shear strains further lowers the  magnetic symmetry group of the system to  $P_s\bar{1}$ MSG (No. 2.7), whose generators include only $\mathcal{T'}$, and inversion $\mathcal{I}$. 

The presence of $\mathcal{T'}$ and $\mathcal{I}$ symmetries enable the possibility to calculate the topological index $\eta_{4I}$ \cite{elcoro2021magnetic}.
Through the employment of  MTQC, the topological invariant $\eta_{4I} = 2$ for the three different cases suggest a nontrivial topology. This indicates that these states can be classified as an axion insulator, potentially leading to the emergence of hinge modes and the half-quantum Hall effect on the side surface states situated at the gap of the surface states on the top and bottom (001) surfaces \cite{gu2021spectral,zhao2021routes}. Furthermore, for both longitudinal shear scenarios, the $k_x = 0$ plane reveals a nontrivial mirror Chern number. This indicates that \ce{EuCd2As2} also qualifies as a topological crystalline insulator, characterized by gapless Dirac surface states on the (010) surface, which are protected by this nonzero mirror Chern number.

\subsection{Shear strain effect on the \textit{afm-ip} \texorpdfstring{$EuCd2As2$} state}
\label{sectionIIIb}

Without any lattice deformation, the \textit{afm-ip} order of \ce{EuCd2As2} is an axion insulator with a nontrivial topological index $\eta_{4I} = 2$. The two bands near the Fermi level (Fig. \ref{ip-strain}a first column) belong to the same irreducible representation \cite{valadkhani2023influence},  resulting in a bulk gap of approximately $10$ meV at the Dirac point. This main gap opening is due the breaking of the  $C_3$-fold rotation symmetry as a consequence of the in-plane N\'eel vector orientation. 
The corresponding magnetic group is $\mathrm{C_c}2/m$  (No. 12.63) with  $\mathcal{T'}$, inversion $\mathcal{I}$, and a two-fold rotation $C_{2x}$ as the generators. The surface states on the (001) (inset Fig.\ref{ip-strain}(b) first column) and (010) surfaces are gapless  and protected by the mirror $M_x = \mathcal{I}C_{2x}$ symmetry due to nontrivial mirror Chern numbers on the $k_x = 0$ plane. This implies that \ce{EuCd2As2} 
with \textit{afm-ip} order is a topological crystalline insulator. Actually, \ce{EuCd2As2} is  a higher order topological insulator exhibiting a gap of 1 meV on the (101) surface, which can be challenging to observe by ARPES measurements \cite{ma2020emergence}. Detecting this phase also would require the examination of  special edges along the (101) surface in order to localize the emerging hinge states.


We again subject the system to shear strain, up to $\pm 3\%$ change in the lattice parameters Fig \ref{ip-strain}. Both longitudinal shear strains do not cause a reduction of the system's magnetic symmetry group, but the diagonal shear strains does, resulting in the $P_s\bar{1}$ MSG (No. 2.7). All the mentioned strains open a gap on the (001) surface and enhance the gap in both the bulk and surface while remarkably preserving the axion insulating phase.

\section{Conclusion}
\label{sectionIV}

In this work we have investigated the effects of in-plane shear strains on the band topology and the surface states of AFM \ce{EuCd2As2}. We have shown the possibility to change the topological features of the AFM \ce{EuCd2As2} by shear strain, which allows us to access three different topological insulators, as well as enhancing the gap on the (001) surface. Besides these new predictions, we have also confirmed previous theoretical studies on this material. 

For a magnetic order pointing out-of-plane, we propose that by applying longitudinal and diagonal shear strains within a 3\% lattice parameter modulation the \ce{EuCd2As2} transforms to an axion insulator phase. Its non-trivial topology is characterized by the parity based invariant $\eta_{4I} = 2$. Additionally, we observe gapped surface states up to a 13 meV on the (001) surfaces, which can be favorable for experimental detection and the possibility to observe hinge modes along those edges. On the other hand, for in-plane magnetic order, the magnetic symmetry group of the unstrained structure is lowered to $\mathrm{C_c}2/m$ and the (001) surface states are gapped. This topology remains unaltered for all shear strains.  Finally, in the presence of a mirror (glide mirror) symmetry, \ce{EuCd2As2} is classified as an topological crystalline insulator with protected gapless surface states emerging on the (010) surfaces.

\begin{acknowledgements}
We acknowledge funding by the Deutsche Forschungsgemeinschaft (DFG, German Research Foundation) — TRR 288 – 422213477 (project B05 and A09). LS acknowledge support by Grant Agency of the Czech Republic grant no. 19-28375X.
\end{acknowledgements}
\bibliographystyle{apsrev4-2}
\bibliography{references}

\begin{thebibliography}{29}%
\makeatletter
\providecommand \@ifxundefined [1]{%
 \@ifx{#1\undefined}
}%
\providecommand \@ifnum [1]{%
 \ifnum #1\expandafter \@firstoftwo
 \else \expandafter \@secondoftwo
 \fi
}%
\providecommand \@ifx [1]{%
 \ifx #1\expandafter \@firstoftwo
 \else \expandafter \@secondoftwo
 \fi
}%
\providecommand \natexlab [1]{#1}%
\providecommand \enquote  [1]{``#1''}%
\providecommand \bibnamefont  [1]{#1}%
\providecommand \bibfnamefont [1]{#1}%
\providecommand \citenamefont [1]{#1}%
\providecommand \href@noop [0]{\@secondoftwo}%
\providecommand \href [0]{\begingroup \@sanitize@url \@href}%
\providecommand \@href[1]{\@@startlink{#1}\@@href}%
\providecommand \@@href[1]{\endgroup#1\@@endlink}%
\providecommand \@sanitize@url [0]{\catcode `\\12\catcode `\$12\catcode
  `\&12\catcode `\#12\catcode `\^12\catcode `\_12\catcode `\%12\relax}%
\providecommand \@@startlink[1]{}%
\providecommand \@@endlink[0]{}%
\providecommand \url  [0]{\begingroup\@sanitize@url \@url }%
\providecommand \@url [1]{\endgroup\@href {#1}{\urlprefix }}%
\providecommand \urlprefix  [0]{URL }%
\providecommand \Eprint [0]{\href }%
\providecommand \doibase [0]{https://doi.org/}%
\providecommand \selectlanguage [0]{\@gobble}%
\providecommand \bibinfo  [0]{\@secondoftwo}%
\providecommand \bibfield  [0]{\@secondoftwo}%
\providecommand \translation [1]{[#1]}%
\providecommand \BibitemOpen [0]{}%
\providecommand \bibitemStop [0]{}%
\providecommand \bibitemNoStop [0]{.\EOS\space}%
\providecommand \EOS [0]{\spacefactor3000\relax}%
\providecommand \BibitemShut  [1]{\csname bibitem#1\endcsname}%
\let\auto@bib@innerbib\@empty
\bibitem [{\citenamefont {Xu}\ \emph {et~al.}(2020)\citenamefont {Xu},
  \citenamefont {Elcoro}, \citenamefont {Song}, \citenamefont {Wieder},
  \citenamefont {Vergniory}, \citenamefont {Regnault}, \citenamefont {Chen},
  \citenamefont {Felser},\ and\ \citenamefont {Bernevig}}]{xu2020high}%
  \BibitemOpen
  \bibfield  {author} {\bibinfo {author} {\bibfnamefont {Y.}~\bibnamefont
  {Xu}}, \bibinfo {author} {\bibfnamefont {L.}~\bibnamefont {Elcoro}}, \bibinfo
  {author} {\bibfnamefont {Z.-D.}\ \bibnamefont {Song}}, \bibinfo {author}
  {\bibfnamefont {B.~J.}\ \bibnamefont {Wieder}}, \bibinfo {author}
  {\bibfnamefont {M.}~\bibnamefont {Vergniory}}, \bibinfo {author}
  {\bibfnamefont {N.}~\bibnamefont {Regnault}}, \bibinfo {author}
  {\bibfnamefont {Y.}~\bibnamefont {Chen}}, \bibinfo {author} {\bibfnamefont
  {C.}~\bibnamefont {Felser}},\ and\ \bibinfo {author} {\bibfnamefont {B.~A.}\
  \bibnamefont {Bernevig}},\ }\href@noop {} {\bibfield  {journal} {\bibinfo
  {journal} {Nature}\ }\textbf {\bibinfo {volume} {586}},\ \bibinfo {pages}
  {702} (\bibinfo {year} {2020})}\BibitemShut {NoStop}%
\bibitem [{\citenamefont {Otrokov}\ \emph {et~al.}(2019)\citenamefont
  {Otrokov}, \citenamefont {Klimovskikh}, \citenamefont {Bentmann},
  \citenamefont {Estyunin}, \citenamefont {Zeugner}, \citenamefont {Aliev},
  \citenamefont {Ga{\ss}}, \citenamefont {Wolter}, \citenamefont {Koroleva},
  \citenamefont {Shikin} \emph {et~al.}}]{otrokov2019prediction}%
  \BibitemOpen
  \bibfield  {author} {\bibinfo {author} {\bibfnamefont {M.~M.}\ \bibnamefont
  {Otrokov}}, \bibinfo {author} {\bibfnamefont {I.~I.}\ \bibnamefont
  {Klimovskikh}}, \bibinfo {author} {\bibfnamefont {H.}~\bibnamefont
  {Bentmann}}, \bibinfo {author} {\bibfnamefont {D.}~\bibnamefont {Estyunin}},
  \bibinfo {author} {\bibfnamefont {A.}~\bibnamefont {Zeugner}}, \bibinfo
  {author} {\bibfnamefont {Z.~S.}\ \bibnamefont {Aliev}}, \bibinfo {author}
  {\bibfnamefont {S.}~\bibnamefont {Ga{\ss}}}, \bibinfo {author} {\bibfnamefont
  {A.}~\bibnamefont {Wolter}}, \bibinfo {author} {\bibfnamefont
  {A.}~\bibnamefont {Koroleva}}, \bibinfo {author} {\bibfnamefont {A.~M.}\
  \bibnamefont {Shikin}}, \emph {et~al.},\ }\href@noop {} {\bibfield  {journal}
  {\bibinfo  {journal} {Nature}\ }\textbf {\bibinfo {volume} {576}},\ \bibinfo
  {pages} {416} (\bibinfo {year} {2019})}\BibitemShut {NoStop}%
\bibitem [{\citenamefont {Li}\ \emph {et~al.}(2019)\citenamefont {Li},
  \citenamefont {Gao}, \citenamefont {Duan}, \citenamefont {Xu}, \citenamefont
  {Zhu}, \citenamefont {Tian}, \citenamefont {Gao}, \citenamefont {Fan},
  \citenamefont {Rao}, \citenamefont {Huang} \emph {et~al.}}]{li2019dirac}%
  \BibitemOpen
  \bibfield  {author} {\bibinfo {author} {\bibfnamefont {H.}~\bibnamefont
  {Li}}, \bibinfo {author} {\bibfnamefont {S.-Y.}\ \bibnamefont {Gao}},
  \bibinfo {author} {\bibfnamefont {S.-F.}\ \bibnamefont {Duan}}, \bibinfo
  {author} {\bibfnamefont {Y.-F.}\ \bibnamefont {Xu}}, \bibinfo {author}
  {\bibfnamefont {K.-J.}\ \bibnamefont {Zhu}}, \bibinfo {author} {\bibfnamefont
  {S.-J.}\ \bibnamefont {Tian}}, \bibinfo {author} {\bibfnamefont {J.-C.}\
  \bibnamefont {Gao}}, \bibinfo {author} {\bibfnamefont {W.-H.}\ \bibnamefont
  {Fan}}, \bibinfo {author} {\bibfnamefont {Z.-C.}\ \bibnamefont {Rao}},
  \bibinfo {author} {\bibfnamefont {J.-R.}\ \bibnamefont {Huang}}, \emph
  {et~al.},\ }\href@noop {} {\bibfield  {journal} {\bibinfo  {journal}
  {Physical Review X}\ }\textbf {\bibinfo {volume} {9}},\ \bibinfo {pages}
  {041039} (\bibinfo {year} {2019})}\BibitemShut {NoStop}%
\bibitem [{\citenamefont {Zhang}\ \emph {et~al.}(2019)\citenamefont {Zhang},
  \citenamefont {Shi}, \citenamefont {Zhu}, \citenamefont {Xing}, \citenamefont
  {Zhang},\ and\ \citenamefont {Wang}}]{zhang2019topological}%
  \BibitemOpen
  \bibfield  {author} {\bibinfo {author} {\bibfnamefont {D.}~\bibnamefont
  {Zhang}}, \bibinfo {author} {\bibfnamefont {M.}~\bibnamefont {Shi}}, \bibinfo
  {author} {\bibfnamefont {T.}~\bibnamefont {Zhu}}, \bibinfo {author}
  {\bibfnamefont {D.}~\bibnamefont {Xing}}, \bibinfo {author} {\bibfnamefont
  {H.}~\bibnamefont {Zhang}},\ and\ \bibinfo {author} {\bibfnamefont
  {J.}~\bibnamefont {Wang}},\ }\href@noop {} {\bibfield  {journal} {\bibinfo
  {journal} {Physical review letters}\ }\textbf {\bibinfo {volume} {122}},\
  \bibinfo {pages} {206401} (\bibinfo {year} {2019})}\BibitemShut {NoStop}%
\bibitem [{\citenamefont {Zou}\ \emph {et~al.}(2019)\citenamefont {Zou},
  \citenamefont {He},\ and\ \citenamefont {Xu}}]{zou2019study}%
  \BibitemOpen
  \bibfield  {author} {\bibinfo {author} {\bibfnamefont {J.}~\bibnamefont
  {Zou}}, \bibinfo {author} {\bibfnamefont {Z.}~\bibnamefont {He}},\ and\
  \bibinfo {author} {\bibfnamefont {G.}~\bibnamefont {Xu}},\ }\href@noop {}
  {\bibfield  {journal} {\bibinfo  {journal} {npj Computational Materials}\
  }\textbf {\bibinfo {volume} {5}},\ \bibinfo {pages} {96} (\bibinfo {year}
  {2019})}\BibitemShut {NoStop}%
\bibitem [{\citenamefont {Hua}\ \emph {et~al.}(2018)\citenamefont {Hua},
  \citenamefont {Nie}, \citenamefont {Song}, \citenamefont {Yu}, \citenamefont
  {Xu},\ and\ \citenamefont {Yao}}]{hua2018dirac}%
  \BibitemOpen
  \bibfield  {author} {\bibinfo {author} {\bibfnamefont {G.}~\bibnamefont
  {Hua}}, \bibinfo {author} {\bibfnamefont {S.}~\bibnamefont {Nie}}, \bibinfo
  {author} {\bibfnamefont {Z.}~\bibnamefont {Song}}, \bibinfo {author}
  {\bibfnamefont {R.}~\bibnamefont {Yu}}, \bibinfo {author} {\bibfnamefont
  {G.}~\bibnamefont {Xu}},\ and\ \bibinfo {author} {\bibfnamefont
  {K.}~\bibnamefont {Yao}},\ }\href@noop {} {\bibfield  {journal} {\bibinfo
  {journal} {Physical Review B}\ }\textbf {\bibinfo {volume} {98}},\ \bibinfo
  {pages} {201116} (\bibinfo {year} {2018})}\BibitemShut {NoStop}%
\bibitem [{\citenamefont {Bernevig}\ \emph {et~al.}(2022)\citenamefont
  {Bernevig}, \citenamefont {Felser},\ and\ \citenamefont
  {Beidenkopf}}]{bernevig2022progress}%
  \BibitemOpen
  \bibfield  {author} {\bibinfo {author} {\bibfnamefont {B.~A.}\ \bibnamefont
  {Bernevig}}, \bibinfo {author} {\bibfnamefont {C.}~\bibnamefont {Felser}},\
  and\ \bibinfo {author} {\bibfnamefont {H.}~\bibnamefont {Beidenkopf}},\
  }\href@noop {} {\bibfield  {journal} {\bibinfo  {journal} {Nature}\ }\textbf
  {\bibinfo {volume} {603}},\ \bibinfo {pages} {41} (\bibinfo {year}
  {2022})}\BibitemShut {NoStop}%
\bibitem [{\citenamefont {{\v{S}}mejkal}\ \emph {et~al.}(2018)\citenamefont
  {{\v{S}}mejkal}, \citenamefont {Mokrousov}, \citenamefont {Yan},\ and\
  \citenamefont {MacDonald}}]{vsmejkal2018topological}%
  \BibitemOpen
  \bibfield  {author} {\bibinfo {author} {\bibfnamefont {L.}~\bibnamefont
  {{\v{S}}mejkal}}, \bibinfo {author} {\bibfnamefont {Y.}~\bibnamefont
  {Mokrousov}}, \bibinfo {author} {\bibfnamefont {B.}~\bibnamefont {Yan}},\
  and\ \bibinfo {author} {\bibfnamefont {A.~H.}\ \bibnamefont {MacDonald}},\
  }\href@noop {} {\bibfield  {journal} {\bibinfo  {journal} {Nature physics}\
  }\textbf {\bibinfo {volume} {14}},\ \bibinfo {pages} {242} (\bibinfo {year}
  {2018})}\BibitemShut {NoStop}%
\bibitem [{\citenamefont {Fan}\ and\ \citenamefont
  {Wang}(2016)}]{fan2016spintronics}%
  \BibitemOpen
  \bibfield  {author} {\bibinfo {author} {\bibfnamefont {Y.}~\bibnamefont
  {Fan}}\ and\ \bibinfo {author} {\bibfnamefont {K.~L.}\ \bibnamefont {Wang}},\
  }in\ \href@noop {} {\emph {\bibinfo {booktitle} {Spin}}},\ Vol.~\bibinfo
  {volume} {6}\ (\bibinfo {organization} {World Scientific},\ \bibinfo {year}
  {2016})\ p.\ \bibinfo {pages} {1640001}\BibitemShut {NoStop}%
\bibitem [{\citenamefont {Zheng}\ \emph {et~al.}(2017)\citenamefont {Zheng},
  \citenamefont {Zhu}, \citenamefont {Liu}, \citenamefont {Lu}, \citenamefont
  {Ning}, \citenamefont {Zhang}, \citenamefont {Gao}, \citenamefont {Han},
  \citenamefont {Yang}, \citenamefont {Du} \emph {et~al.}}]{zheng2017field}%
  \BibitemOpen
  \bibfield  {author} {\bibinfo {author} {\bibfnamefont {G.}~\bibnamefont
  {Zheng}}, \bibinfo {author} {\bibfnamefont {X.}~\bibnamefont {Zhu}}, \bibinfo
  {author} {\bibfnamefont {Y.}~\bibnamefont {Liu}}, \bibinfo {author}
  {\bibfnamefont {J.}~\bibnamefont {Lu}}, \bibinfo {author} {\bibfnamefont
  {W.}~\bibnamefont {Ning}}, \bibinfo {author} {\bibfnamefont {H.}~\bibnamefont
  {Zhang}}, \bibinfo {author} {\bibfnamefont {W.}~\bibnamefont {Gao}}, \bibinfo
  {author} {\bibfnamefont {Y.}~\bibnamefont {Han}}, \bibinfo {author}
  {\bibfnamefont {J.}~\bibnamefont {Yang}}, \bibinfo {author} {\bibfnamefont
  {H.}~\bibnamefont {Du}}, \emph {et~al.},\ }\href@noop {} {\bibfield
  {journal} {\bibinfo  {journal} {Physical Review B}\ }\textbf {\bibinfo
  {volume} {96}},\ \bibinfo {pages} {121401} (\bibinfo {year}
  {2017})}\BibitemShut {NoStop}%
\bibitem [{\citenamefont {Li}\ \emph {et~al.}(2020)\citenamefont {Li},
  \citenamefont {Jiang}, \citenamefont {Zhang}, \citenamefont {Liu},
  \citenamefont {Yang},\ and\ \citenamefont {Wang}}]{li2020intrinsic}%
  \BibitemOpen
  \bibfield  {author} {\bibinfo {author} {\bibfnamefont {Y.}~\bibnamefont
  {Li}}, \bibinfo {author} {\bibfnamefont {Y.}~\bibnamefont {Jiang}}, \bibinfo
  {author} {\bibfnamefont {J.}~\bibnamefont {Zhang}}, \bibinfo {author}
  {\bibfnamefont {Z.}~\bibnamefont {Liu}}, \bibinfo {author} {\bibfnamefont
  {Z.}~\bibnamefont {Yang}},\ and\ \bibinfo {author} {\bibfnamefont
  {J.}~\bibnamefont {Wang}},\ }\href@noop {} {\bibfield  {journal} {\bibinfo
  {journal} {Physical Review B}\ }\textbf {\bibinfo {volume} {102}},\ \bibinfo
  {pages} {121107} (\bibinfo {year} {2020})}\BibitemShut {NoStop}%
\bibitem [{\citenamefont {Zheng}\ \emph {et~al.}(2015)\citenamefont {Zheng},
  \citenamefont {Shen}, \citenamefont {Wang},\ and\ \citenamefont
  {Zhai}}]{zheng2015magnetic}%
  \BibitemOpen
  \bibfield  {author} {\bibinfo {author} {\bibfnamefont {W.}~\bibnamefont
  {Zheng}}, \bibinfo {author} {\bibfnamefont {H.}~\bibnamefont {Shen}},
  \bibinfo {author} {\bibfnamefont {Z.}~\bibnamefont {Wang}},\ and\ \bibinfo
  {author} {\bibfnamefont {H.}~\bibnamefont {Zhai}},\ }\href@noop {} {\bibfield
   {journal} {\bibinfo  {journal} {Physical Review B}\ }\textbf {\bibinfo
  {volume} {91}},\ \bibinfo {pages} {161107} (\bibinfo {year}
  {2015})}\BibitemShut {NoStop}%
\bibitem [{\citenamefont {Liu}\ \emph {et~al.}(2014{\natexlab{a}})\citenamefont
  {Liu}, \citenamefont {Jiang}, \citenamefont {Zhou}, \citenamefont {Wang},
  \citenamefont {Zhang}, \citenamefont {Weng}, \citenamefont {Prabhakaran},
  \citenamefont {Mo}, \citenamefont {Peng}, \citenamefont {Dudin} \emph
  {et~al.}}]{liu2014stable}%
  \BibitemOpen
  \bibfield  {author} {\bibinfo {author} {\bibfnamefont {Z.}~\bibnamefont
  {Liu}}, \bibinfo {author} {\bibfnamefont {J.}~\bibnamefont {Jiang}}, \bibinfo
  {author} {\bibfnamefont {B.}~\bibnamefont {Zhou}}, \bibinfo {author}
  {\bibfnamefont {Z.}~\bibnamefont {Wang}}, \bibinfo {author} {\bibfnamefont
  {Y.}~\bibnamefont {Zhang}}, \bibinfo {author} {\bibfnamefont
  {H.}~\bibnamefont {Weng}}, \bibinfo {author} {\bibfnamefont {D.}~\bibnamefont
  {Prabhakaran}}, \bibinfo {author} {\bibfnamefont {S.~K.}\ \bibnamefont {Mo}},
  \bibinfo {author} {\bibfnamefont {H.}~\bibnamefont {Peng}}, \bibinfo {author}
  {\bibfnamefont {P.}~\bibnamefont {Dudin}}, \emph {et~al.},\ }\href@noop {}
  {\bibfield  {journal} {\bibinfo  {journal} {Nature materials}\ }\textbf
  {\bibinfo {volume} {13}},\ \bibinfo {pages} {677} (\bibinfo {year}
  {2014}{\natexlab{a}})}\BibitemShut {NoStop}%
\bibitem [{\citenamefont {Aggarwal}\ \emph {et~al.}(2016)\citenamefont
  {Aggarwal}, \citenamefont {Gaurav}, \citenamefont {Thakur}, \citenamefont
  {Haque}, \citenamefont {Ganguli},\ and\ \citenamefont
  {Sheet}}]{aggarwal2016unconventional}%
  \BibitemOpen
  \bibfield  {author} {\bibinfo {author} {\bibfnamefont {L.}~\bibnamefont
  {Aggarwal}}, \bibinfo {author} {\bibfnamefont {A.}~\bibnamefont {Gaurav}},
  \bibinfo {author} {\bibfnamefont {G.~S.}\ \bibnamefont {Thakur}}, \bibinfo
  {author} {\bibfnamefont {Z.}~\bibnamefont {Haque}}, \bibinfo {author}
  {\bibfnamefont {A.~K.}\ \bibnamefont {Ganguli}},\ and\ \bibinfo {author}
  {\bibfnamefont {G.}~\bibnamefont {Sheet}},\ }\href@noop {} {\bibfield
  {journal} {\bibinfo  {journal} {Nature materials}\ }\textbf {\bibinfo
  {volume} {15}},\ \bibinfo {pages} {32} (\bibinfo {year} {2016})}\BibitemShut
  {NoStop}%
\bibitem [{\citenamefont {Neupane}\ \emph {et~al.}(2014)\citenamefont
  {Neupane}, \citenamefont {Xu}, \citenamefont {Sankar}, \citenamefont
  {Alidoust}, \citenamefont {Bian}, \citenamefont {Liu}, \citenamefont
  {Belopolski}, \citenamefont {Chang}, \citenamefont {Jeng}, \citenamefont
  {Lin} \emph {et~al.}}]{neupane2014observation}%
  \BibitemOpen
  \bibfield  {author} {\bibinfo {author} {\bibfnamefont {M.}~\bibnamefont
  {Neupane}}, \bibinfo {author} {\bibfnamefont {S.-Y.}\ \bibnamefont {Xu}},
  \bibinfo {author} {\bibfnamefont {R.}~\bibnamefont {Sankar}}, \bibinfo
  {author} {\bibfnamefont {N.}~\bibnamefont {Alidoust}}, \bibinfo {author}
  {\bibfnamefont {G.}~\bibnamefont {Bian}}, \bibinfo {author} {\bibfnamefont
  {C.}~\bibnamefont {Liu}}, \bibinfo {author} {\bibfnamefont {I.}~\bibnamefont
  {Belopolski}}, \bibinfo {author} {\bibfnamefont {T.-R.}\ \bibnamefont
  {Chang}}, \bibinfo {author} {\bibfnamefont {H.-T.}\ \bibnamefont {Jeng}},
  \bibinfo {author} {\bibfnamefont {H.}~\bibnamefont {Lin}}, \emph {et~al.},\
  }\href@noop {} {\bibfield  {journal} {\bibinfo  {journal} {Nature
  communications}\ }\textbf {\bibinfo {volume} {5}},\ \bibinfo {pages} {3786}
  (\bibinfo {year} {2014})}\BibitemShut {NoStop}%
\bibitem [{\citenamefont {Ma}\ \emph {et~al.}(2020)\citenamefont {Ma},
  \citenamefont {Wang}, \citenamefont {Nie}, \citenamefont {Yi}, \citenamefont
  {Xu}, \citenamefont {Li}, \citenamefont {Jandke}, \citenamefont {Wulfhekel},
  \citenamefont {Huang}, \citenamefont {West} \emph
  {et~al.}}]{ma2020emergence}%
  \BibitemOpen
  \bibfield  {author} {\bibinfo {author} {\bibfnamefont {J.}~\bibnamefont
  {Ma}}, \bibinfo {author} {\bibfnamefont {H.}~\bibnamefont {Wang}}, \bibinfo
  {author} {\bibfnamefont {S.}~\bibnamefont {Nie}}, \bibinfo {author}
  {\bibfnamefont {C.}~\bibnamefont {Yi}}, \bibinfo {author} {\bibfnamefont
  {Y.}~\bibnamefont {Xu}}, \bibinfo {author} {\bibfnamefont {H.}~\bibnamefont
  {Li}}, \bibinfo {author} {\bibfnamefont {J.}~\bibnamefont {Jandke}}, \bibinfo
  {author} {\bibfnamefont {W.}~\bibnamefont {Wulfhekel}}, \bibinfo {author}
  {\bibfnamefont {Y.}~\bibnamefont {Huang}}, \bibinfo {author} {\bibfnamefont
  {D.}~\bibnamefont {West}}, \emph {et~al.},\ }\href@noop {} {\bibfield
  {journal} {\bibinfo  {journal} {Advanced Materials}\ }\textbf {\bibinfo
  {volume} {32}},\ \bibinfo {pages} {1907565} (\bibinfo {year}
  {2020})}\BibitemShut {NoStop}%
\bibitem [{\citenamefont {Cuono}\ \emph {et~al.}(2023)\citenamefont {Cuono},
  \citenamefont {Sattigeri}, \citenamefont {Autieri},\ and\ \citenamefont
  {Dietl}}]{cuono2023}%
  \BibitemOpen
  \bibfield  {author} {\bibinfo {author} {\bibfnamefont {G.}~\bibnamefont
  {Cuono}}, \bibinfo {author} {\bibfnamefont {R.~M.}\ \bibnamefont
  {Sattigeri}}, \bibinfo {author} {\bibfnamefont {C.}~\bibnamefont {Autieri}},\
  and\ \bibinfo {author} {\bibfnamefont {T.}~\bibnamefont {Dietl}},\
  }\href@noop {} {\bibfield  {journal} {\bibinfo  {journal} {Physical Review
  B}\ }\textbf {\bibinfo {volume} {108}},\ \bibinfo {pages} {075150} (\bibinfo
  {year} {2023})}\BibitemShut {NoStop}%
\bibitem [{\citenamefont {Rahn}\ \emph {et~al.}(2018)\citenamefont {Rahn},
  \citenamefont {Soh}, \citenamefont {Francoual}, \citenamefont {Veiga},
  \citenamefont {Strempfer}, \citenamefont {Mardegan}, \citenamefont {Yan},
  \citenamefont {Guo}, \citenamefont {Shi},\ and\ \citenamefont
  {Boothroyd}}]{rahn2018coupling}%
  \BibitemOpen
  \bibfield  {author} {\bibinfo {author} {\bibfnamefont {M.}~\bibnamefont
  {Rahn}}, \bibinfo {author} {\bibfnamefont {J.-R.}\ \bibnamefont {Soh}},
  \bibinfo {author} {\bibfnamefont {S.}~\bibnamefont {Francoual}}, \bibinfo
  {author} {\bibfnamefont {L.}~\bibnamefont {Veiga}}, \bibinfo {author}
  {\bibfnamefont {J.}~\bibnamefont {Strempfer}}, \bibinfo {author}
  {\bibfnamefont {J.}~\bibnamefont {Mardegan}}, \bibinfo {author}
  {\bibfnamefont {D.}~\bibnamefont {Yan}}, \bibinfo {author} {\bibfnamefont
  {Y.}~\bibnamefont {Guo}}, \bibinfo {author} {\bibfnamefont {Y.}~\bibnamefont
  {Shi}},\ and\ \bibinfo {author} {\bibfnamefont {A.}~\bibnamefont
  {Boothroyd}},\ }\href@noop {} {\bibfield  {journal} {\bibinfo  {journal}
  {Physical Review B}\ }\textbf {\bibinfo {volume} {97}},\ \bibinfo {pages}
  {214422} (\bibinfo {year} {2018})}\BibitemShut {NoStop}%
\bibitem [{\citenamefont {Valadkhani}\ \emph {et~al.}(2023)\citenamefont
  {Valadkhani}, \citenamefont {Iraola}, \citenamefont {F{\"u}nfhaus},
  \citenamefont {Song}, \citenamefont {Smejkal}, \citenamefont {Sinova},\ and\
  \citenamefont {Valenti}}]{valadkhani2023influence}%
  \BibitemOpen
  \bibfield  {author} {\bibinfo {author} {\bibfnamefont {A.}~\bibnamefont
  {Valadkhani}}, \bibinfo {author} {\bibfnamefont {M.}~\bibnamefont {Iraola}},
  \bibinfo {author} {\bibfnamefont {A.}~\bibnamefont {F{\"u}nfhaus}}, \bibinfo
  {author} {\bibfnamefont {Y.-J.}\ \bibnamefont {Song}}, \bibinfo {author}
  {\bibfnamefont {L.}~\bibnamefont {Smejkal}}, \bibinfo {author} {\bibfnamefont
  {J.}~\bibnamefont {Sinova}},\ and\ \bibinfo {author} {\bibfnamefont
  {R.}~\bibnamefont {Valenti}},\ }\href@noop {} {\bibfield  {journal} {\bibinfo
   {journal} {arXiv preprint arXiv:2308.08619}\ } (\bibinfo {year}
  {2023})}\BibitemShut {NoStop}%
\bibitem [{\citenamefont {Hafner}(2008)}]{hafner2008ab}%
  \BibitemOpen
  \bibfield  {author} {\bibinfo {author} {\bibfnamefont {J.}~\bibnamefont
  {Hafner}},\ }\href@noop {} {\bibfield  {journal} {\bibinfo  {journal}
  {Journal of computational chemistry}\ }\textbf {\bibinfo {volume} {29}},\
  \bibinfo {pages} {2044} (\bibinfo {year} {2008})}\BibitemShut {NoStop}%
\bibitem [{\citenamefont {Perdew}\ \emph {et~al.}(1998)\citenamefont {Perdew},
  \citenamefont {Burke},\ and\ \citenamefont {Ernzerhof}}]{perdew1998perdew}%
  \BibitemOpen
  \bibfield  {author} {\bibinfo {author} {\bibfnamefont {J.~P.}\ \bibnamefont
  {Perdew}}, \bibinfo {author} {\bibfnamefont {K.}~\bibnamefont {Burke}},\ and\
  \bibinfo {author} {\bibfnamefont {M.}~\bibnamefont {Ernzerhof}},\ }\href@noop
  {} {\bibfield  {journal} {\bibinfo  {journal} {Physical Review Letters}\
  }\textbf {\bibinfo {volume} {80}},\ \bibinfo {pages} {891} (\bibinfo {year}
  {1998})}\BibitemShut {NoStop}%
\bibitem [{\citenamefont {Schellenberg}\ \emph {et~al.}(2011)\citenamefont
  {Schellenberg}, \citenamefont {Pfannenschmidt}, \citenamefont {Eul},
  \citenamefont {Schwickert},\ and\ \citenamefont
  {P{\"o}ttgen}}]{schellenberg2011121sb}%
  \BibitemOpen
  \bibfield  {author} {\bibinfo {author} {\bibfnamefont {I.}~\bibnamefont
  {Schellenberg}}, \bibinfo {author} {\bibfnamefont {U.}~\bibnamefont
  {Pfannenschmidt}}, \bibinfo {author} {\bibfnamefont {M.}~\bibnamefont {Eul}},
  \bibinfo {author} {\bibfnamefont {C.}~\bibnamefont {Schwickert}},\ and\
  \bibinfo {author} {\bibfnamefont {R.}~\bibnamefont {P{\"o}ttgen}},\
  }\href@noop {} {\bibfield  {journal} {\bibinfo  {journal} {Zeitschrift
  f{\"u}r anorganische und allgemeine Chemie}\ }\textbf {\bibinfo {volume}
  {637}},\ \bibinfo {pages} {1863} (\bibinfo {year} {2011})}\BibitemShut
  {NoStop}%
\bibitem [{\citenamefont {Elcoro}\ \emph {et~al.}(2021)\citenamefont {Elcoro},
  \citenamefont {Wieder}, \citenamefont {Song}, \citenamefont {Xu},
  \citenamefont {Bradlyn},\ and\ \citenamefont
  {Bernevig}}]{elcoro2021magnetic}%
  \BibitemOpen
  \bibfield  {author} {\bibinfo {author} {\bibfnamefont {L.}~\bibnamefont
  {Elcoro}}, \bibinfo {author} {\bibfnamefont {B.~J.}\ \bibnamefont {Wieder}},
  \bibinfo {author} {\bibfnamefont {Z.}~\bibnamefont {Song}}, \bibinfo {author}
  {\bibfnamefont {Y.}~\bibnamefont {Xu}}, \bibinfo {author} {\bibfnamefont
  {B.}~\bibnamefont {Bradlyn}},\ and\ \bibinfo {author} {\bibfnamefont {B.~A.}\
  \bibnamefont {Bernevig}},\ }\href@noop {} {\bibfield  {journal} {\bibinfo
  {journal} {Nature communications}\ }\textbf {\bibinfo {volume} {12}},\
  \bibinfo {pages} {5965} (\bibinfo {year} {2021})}\BibitemShut {NoStop}%
\bibitem [{\citenamefont {Vergniory}\ \emph {et~al.}(2019)\citenamefont
  {Vergniory}, \citenamefont {Elcoro}, \citenamefont {Felser}, \citenamefont
  {Regnault}, \citenamefont {Bernevig},\ and\ \citenamefont
  {Wang}}]{vergniory2019complete}%
  \BibitemOpen
  \bibfield  {author} {\bibinfo {author} {\bibfnamefont {M.}~\bibnamefont
  {Vergniory}}, \bibinfo {author} {\bibfnamefont {L.}~\bibnamefont {Elcoro}},
  \bibinfo {author} {\bibfnamefont {C.}~\bibnamefont {Felser}}, \bibinfo
  {author} {\bibfnamefont {N.}~\bibnamefont {Regnault}}, \bibinfo {author}
  {\bibfnamefont {B.~A.}\ \bibnamefont {Bernevig}},\ and\ \bibinfo {author}
  {\bibfnamefont {Z.}~\bibnamefont {Wang}},\ }\href@noop {} {\bibfield
  {journal} {\bibinfo  {journal} {Nature}\ }\textbf {\bibinfo {volume} {566}},\
  \bibinfo {pages} {480} (\bibinfo {year} {2019})}\BibitemShut {NoStop}%
\bibitem [{\citenamefont {Zhi}\ \emph {et~al.}(2022)\citenamefont {Zhi},
  \citenamefont {Xu}, \citenamefont {Wu}, \citenamefont {Ning},\ and\
  \citenamefont {Cao}}]{zhi2022wannsymm}%
  \BibitemOpen
  \bibfield  {author} {\bibinfo {author} {\bibfnamefont {G.-X.}\ \bibnamefont
  {Zhi}}, \bibinfo {author} {\bibfnamefont {C.}~\bibnamefont {Xu}}, \bibinfo
  {author} {\bibfnamefont {S.-Q.}\ \bibnamefont {Wu}}, \bibinfo {author}
  {\bibfnamefont {F.}~\bibnamefont {Ning}},\ and\ \bibinfo {author}
  {\bibfnamefont {C.}~\bibnamefont {Cao}},\ }\href@noop {} {\bibfield
  {journal} {\bibinfo  {journal} {Computer Physics Communications}\ }\textbf
  {\bibinfo {volume} {271}},\ \bibinfo {pages} {108196} (\bibinfo {year}
  {2022})}\BibitemShut {NoStop}%
\bibitem [{\citenamefont {Wu}\ \emph {et~al.}(2018)\citenamefont {Wu},
  \citenamefont {Zhang}, \citenamefont {Song}, \citenamefont {Troyer},\ and\
  \citenamefont {Soluyanov}}]{wu2018wanniertools}%
  \BibitemOpen
  \bibfield  {author} {\bibinfo {author} {\bibfnamefont {Q.}~\bibnamefont
  {Wu}}, \bibinfo {author} {\bibfnamefont {S.}~\bibnamefont {Zhang}}, \bibinfo
  {author} {\bibfnamefont {H.-F.}\ \bibnamefont {Song}}, \bibinfo {author}
  {\bibfnamefont {M.}~\bibnamefont {Troyer}},\ and\ \bibinfo {author}
  {\bibfnamefont {A.~A.}\ \bibnamefont {Soluyanov}},\ }\href@noop {} {\bibfield
   {journal} {\bibinfo  {journal} {Computer Physics Communications}\ }\textbf
  {\bibinfo {volume} {224}},\ \bibinfo {pages} {405} (\bibinfo {year}
  {2018})}\BibitemShut {NoStop}%
\bibitem [{\citenamefont {Liu}\ \emph {et~al.}(2014{\natexlab{b}})\citenamefont
  {Liu}, \citenamefont {Zhou}, \citenamefont {Zhang}, \citenamefont {Wang},
  \citenamefont {Weng}, \citenamefont {Prabhakaran}, \citenamefont {Mo},
  \citenamefont {Shen}, \citenamefont {Fang}, \citenamefont {Dai} \emph
  {et~al.}}]{liu2014discovery}%
  \BibitemOpen
  \bibfield  {author} {\bibinfo {author} {\bibfnamefont {Z.}~\bibnamefont
  {Liu}}, \bibinfo {author} {\bibfnamefont {B.}~\bibnamefont {Zhou}}, \bibinfo
  {author} {\bibfnamefont {Y.}~\bibnamefont {Zhang}}, \bibinfo {author}
  {\bibfnamefont {Z.}~\bibnamefont {Wang}}, \bibinfo {author} {\bibfnamefont
  {H.}~\bibnamefont {Weng}}, \bibinfo {author} {\bibfnamefont {D.}~\bibnamefont
  {Prabhakaran}}, \bibinfo {author} {\bibfnamefont {S.-K.}\ \bibnamefont {Mo}},
  \bibinfo {author} {\bibfnamefont {Z.}~\bibnamefont {Shen}}, \bibinfo {author}
  {\bibfnamefont {Z.}~\bibnamefont {Fang}}, \bibinfo {author} {\bibfnamefont
  {X.}~\bibnamefont {Dai}}, \emph {et~al.},\ }\href@noop {} {\bibfield
  {journal} {\bibinfo  {journal} {Science}\ }\textbf {\bibinfo {volume}
  {343}},\ \bibinfo {pages} {864} (\bibinfo {year}
  {2014}{\natexlab{b}})}\BibitemShut {NoStop}%
\bibitem [{\citenamefont {Gu}\ \emph {et~al.}(2021)\citenamefont {Gu},
  \citenamefont {Li}, \citenamefont {Sun}, \citenamefont {Zhao}, \citenamefont
  {Liu}, \citenamefont {Liu}, \citenamefont {Lu},\ and\ \citenamefont
  {Liu}}]{gu2021spectral}%
  \BibitemOpen
  \bibfield  {author} {\bibinfo {author} {\bibfnamefont {M.}~\bibnamefont
  {Gu}}, \bibinfo {author} {\bibfnamefont {J.}~\bibnamefont {Li}}, \bibinfo
  {author} {\bibfnamefont {H.}~\bibnamefont {Sun}}, \bibinfo {author}
  {\bibfnamefont {Y.}~\bibnamefont {Zhao}}, \bibinfo {author} {\bibfnamefont
  {C.}~\bibnamefont {Liu}}, \bibinfo {author} {\bibfnamefont {J.}~\bibnamefont
  {Liu}}, \bibinfo {author} {\bibfnamefont {H.}~\bibnamefont {Lu}},\ and\
  \bibinfo {author} {\bibfnamefont {Q.}~\bibnamefont {Liu}},\ }\href@noop {}
  {\bibfield  {journal} {\bibinfo  {journal} {Nature communications}\ }\textbf
  {\bibinfo {volume} {12}},\ \bibinfo {pages} {3524} (\bibinfo {year}
  {2021})}\BibitemShut {NoStop}%
\bibitem [{\citenamefont {Zhao}\ and\ \citenamefont
  {Liu}(2021)}]{zhao2021routes}%
  \BibitemOpen
  \bibfield  {author} {\bibinfo {author} {\bibfnamefont {Y.}~\bibnamefont
  {Zhao}}\ and\ \bibinfo {author} {\bibfnamefont {Q.}~\bibnamefont {Liu}},\
  }\href@noop {} {\bibfield  {journal} {\bibinfo  {journal} {Applied Physics
  Letters}\ }\textbf {\bibinfo {volume} {119}} (\bibinfo {year}
  {2021})}\BibitemShut {NoStop}%
\end{thebibliography}%

\end{document}